\documentclass[aps,prl,showpacs,superscriptaddress,groupedaddress,notitlepage]{revtex4-1}

\usepackage[english]{babel}
\usepackage{amsmath,amssymb}
\usepackage{graphicx}
\usepackage{dcolumn}
\usepackage{bm}
\usepackage{amsmath}
\usepackage[applemac]{inputenc}
\usepackage{mathtools}

\usepackage[table,xcdraw]{xcolor}

\begin{document}

\title[The Forest Fire Model Revisited]{The Forest Fire Model Revisited}

\author{Lorenzo Palmieri$^{1}$} 
\email{l.palmieri16@imperial.ac.uk}
\author{Henrik Jeldtoft Jensen$^{1,2}$}
\email{h.jensen@imperial.ac.uk}

\affiliation{{$^1$}Centre for Complexity Science and Department of Mathematics, Imperial College London, South Kensington Campus, SW7 2AZ, UK}
 \affiliation{{$^2$} Institute of Innovative Research, Tokyo Institute of Technology, 4259, Nagatsuta-cho, Yokohama 226-8502, Japan Japan.}

\begin{abstract}
Since  Self-Organised Criticality (SOC) was introduced in the 1987 both the nature of the self-organisation and of the criticality remains controversial. Recent observations on rain precipitation and on brain activity suggest that real systems may dynamically wander about in the vicinity of criticality rather than tune to a critical point. We use computer simulations to study the  Drossel-Schwable forest-fire model of SOC and find that it exhibits behaviour  similar to  that found for rain and brain activity.  In particular we analyse the residence time for different densities of trees and perform finite size scaling analysis of the with of the distribution of residence times.  We conclude that despite of the long known fact that this model does not exhibit exact scaling and power laws its behaviour may exactly for that be very relevant to real physical systems.
\end{abstract}



\maketitle









\section{Introduction}
Much of the research inspired by Self-organised Criticality took up the  mantle from the paper by Bak, Tang and Wiesenfeld\cite{BTW1987} and studied how various simple dynamical systems may drive themselves into a critical state. Reviews of this report can be found in \cite{Jensen1998,Gunnar_Book,25Years}. We are inspired to return to the discussion concerning the nature of the self tuning to a critical state or to the vicinity of such a  state by the similarity found when analysing the size distribution of rain showers \cite{Peters_Neelin2006} and the bursts of brain activity measured during fMRI scans\cite{Chialvo2012}. Both studies find indications of critical behaviour in terms of approximate power laws and even features reminiscent of peaked, or perhaps diverging, fluctuations or susceptibilities. But both studies also investigated the time spent at different values of the control parameters (water vapour and number of voxels activated above threshold respectively). In both cases the distribution of residence times, the amount of time spent at a certain value of the control parameter, is found to exhibit a broad peak centred about what appears to be a critical value of the control. The immediate interpretation seems to be that even for systems as big as the atmosphere or with as many components as the brain the dynamics consists in some kind of feedback mechanism that is able to bring the system into the vicinity of a critical point. However, the dynamics couples the value of the control parameter to the fluctuations in a manner that moves the system around in this vicinity of the critical point. This is similar to suggestions previously put forward such as \cite{Zapperi1995,Sornette1992}. Intuitively, one may imagine something like the following in the case of precipitation: Nucleation of drops happens a critical value of the vapour content. The vapour in the atmosphere over the ocean gradually builds up towards that value and sometimes over shooting may even occur before nucleation is seeded. When precipitation events occur vapour is removed from the atmosphere and one observe oscillations between subcritical and supercritical region. For some reason the coupling between the driving (vapour formation) and the response (precipitation) produces so large  fluctuations that a precise tuning to the critical value of the control parameter is excluded. 

 It was very early realised that one of the paradigmatic models of SOC, the Drossel-Schabl Forest Fire Model (FFM), doesn't exhibit exact scaling as seen for ordinary equilibrium critical systems\cite{Grassberger1993,Pruessner2002,Grassberger2002} and that this may be related to such feedback dynamics. Hitherto, most people have probably considered the lack of scaling in SOC models (the same complicated scaling is also seen for the original BTW sandpile model (See Chap. 4 in \cite{Gunnar_Book}) as a sign of broken promise in the sense that SOC suggested that the scenario of criticality and scaling seen in equilibrium systems were generic for a broad range of driven systems. Here we reanalyse the FFM with the observation on precipitation and brain activity in mind and we suggest that the  model exhibits behaviour which may indeed be of relevance to the type of approximate, or near, critical dynamics seen in some real systems.  
 
In this paper we simulate the FFM using the efficient algorithm described in \cite{Pruessner2002}. We run the model in the usual mode where the the two probabilities with which empty sites become occupied by trees and the probability that a tree catches fire (from lightning) are prescribed and fixed. We then monitor the density of trees, the time the system spends at a given density, this we call the residence time, and the joint distribution of the sizes of tree clusters and the density of trees on the lattice. We analyse the behaviour as a function of the linear size $L$ of the lattice  and find that the residence times to a very good approximation exhibit a Gaussian peak width a width of peak $\sigma(L)$  about a density $\rho^*(L)$.  We find that for large values of $L$ the average tree density converges like $\rho(L)=\rho_\infty+bL^{-\alpha} $ towards the value $\rho_\infty=0.3976\pm 0.0001$, consistent with previously reported values \cite{Pruessner2002} with the $L$ dependence given by $\alpha=2.033\pm0.037$ (These results are for $\theta=1000$, see below for details). We notice that tis density is far below the critical percolation density of about 0.6 for a square lattice. The width of the peak decays with the linear size of the system essentially as $\sigma(L)= 16.82/L$ also in agreement with the analysis in \cite{Pruessner2002}.  The probability to find the system at a tree density more than one standard deviation away from the average is for all systems sizes about 32\%, so very close to Gaussian behaviour.

It is of course difficult to know how to relate the lattice size of the FFM to the effective number of components of real systems such as the atmosphere or the brain, but if the FFM resembles such systems, the slow narrowing of the peak of residence times and the substantial fluctuations beyond one standard deviation for all system sizes  suggest that SOC dynamics even for large systems generates an exploration of a significant range of control parameters rather than a self-tuning to a specific value or an infinitesimal region about such a value.

\section{Model details}
We consider a two dimensional square lattice of linear extension $L$. In the language of the FFM a lattice site can be in one of three states: It may be empty, contain a tree or it may host a tree on fire. In the  original version of the model\cite{Drossel1992} a random site would be chosen with uniform probability. If the site is empty it will become a tree site with probability $p$, if it contains a tree it will spontaneously be turned into a fire site with probability $f$ (that is stuck by a lightning) and if a tree site is nearest neighbour to a site on fire, the tree will become a fire with probability one. Fire sites are turned into empty sites with probability one. Here we implement this procedure by a slightly different, but very efficient, algorithm used in \cite{Pruessner2002,Clar1994,Grassberger1993,Schenk2000}. The main difference between the two algorithms is that now trees are grown in chunks of $\theta = \frac{p}{f}$ between two lightning attempts, this does not affecting the final statistics \cite{Schenk2000}. Here we recall briefly the implementation of the algorithm in pseudo-code.
\\ \\ FOREVER $\lbrace$
\\ \\REPEAT $\theta$ TIMES $\lbrace$ \\
choose randomly a site $s$; \\
IF( $s$ is empty) THEN $\lbrace$ $s$ becomes occupied $\rbrace $
\\      $\rbrace $
\\ 	\\choose randomly a site $s$; \\ 
IF( $s$ is occupied) THEN $\lbrace$ 
\\ collect statistics;
\\ burn the whole cluster related to $s$ ;
\\ $\rbrace $
\\ \\$\rbrace $ \\
The statistics is collected taking into account the overall density of trees, $\rho$, and the size of the burnt cluster $S$. An exception is made for the data represented in Fig. \ref{Order_para_Suscep}, where the biggest cluster has been removed before the computation of the mean and variance of all the clusters that are present in the system at each burning step.
\section{Results}

\begin{figure}[!h]
\centering
\includegraphics[width=16cm,height=7cm]{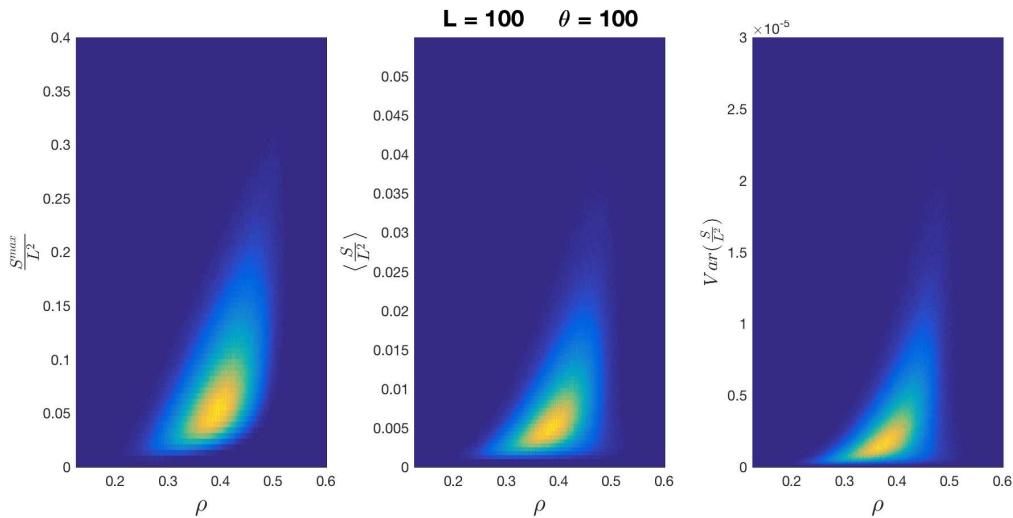}\\
  
\caption{Plots of the max cluster size, its average and the variance. The first two are used as order parameters when analysing ordinary percolation\cite{Christensen_Moloney2005}. The third corresponds to the susceptibility, see e.g. \cite{Ma1985}.  }
\label{Order_para_Suscep}
\end{figure}

We now discuss how the dynamics of the Forest Fire Model is related to the behaviour observed in rain and brain data. See Figs. 1 and 3 in \cite{Peters_Neelin2006}, Figs. 3 panel C, D, E and F in \cite{Chialvo2012} and Fig. 10 in \cite{Scott2014}.
\\We start by looking at the behaviour of the order parameter as function of the density of trees. In the study of rain by Peter and Neelin\cite{Peters_Neelin2006} this corresponds to the precipitation rate as function of water vapour and in human brain studies by Chialvo and collaborators\cite{Chialvo2012} this corresponds to the  normalised size of the the largest activity cluster as function of the total number of active voxels. The analysis of the human brain was later extended to mice, see \cite{Scott2014,Fagerholm2016}, and a phenomenology very similar to the one reported for the human brain was found for mice brains. Inspired by percolation, two choices for the order parameter appear natural. One is the normalised size of the largest cluster of trees, i.e. exactly the quantity used by Chialvo and collaborators in \cite{Chialvo2012} another, equivalent, is the average cluster size. We plot both these in Fig.\ref{Order_para_Suscep} in the form of heat-maps. In the same figure we also plot the variance of the cluster size distribution, this corresponds to the susceptibility in equilibrium statistical mechanics (see e.g. \cite{Ma1985}) which diverges at a critical phase transition. The equivalent variance were studied in both the rain and the brain analysis.  

\begin{figure}[!h]
\centering
\includegraphics[width=16cm,height=7cm]{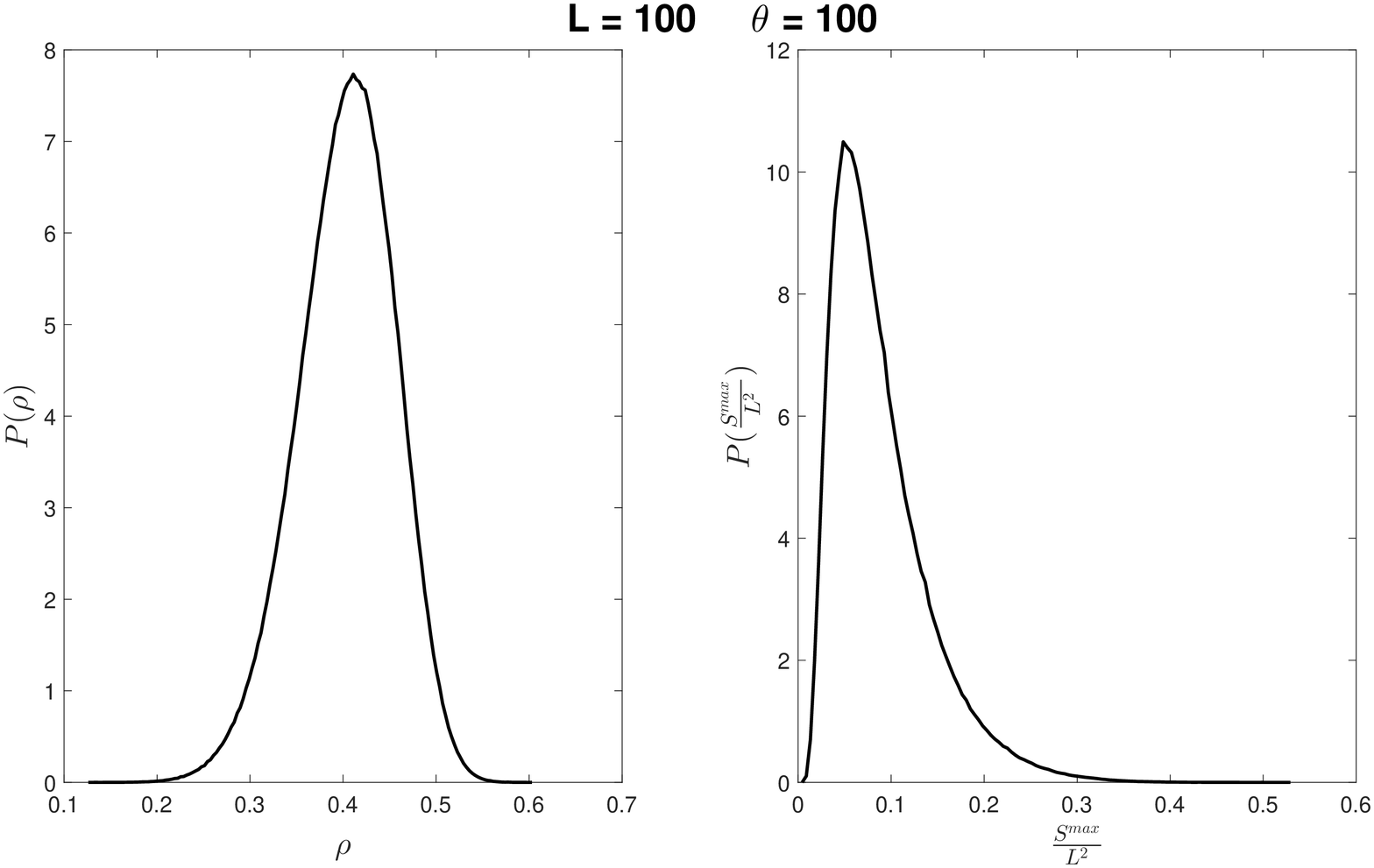}\\
\caption{Left panel contains the residence time distribution for the tree density for $L=100$ and $\theta=100$. The right panel contains the distribution of the order parameter given by the normalised size of the largest cluster for the same values of $L$ and $\theta$.}
\label{Dist_Resi}
\label{pairs}
\end{figure}

To quantify to what extent the dynamics is able to drive the system into a critical state the rain and brain studies both analysed residence times. That is the distribution of times the system is found at a given value of the control parameter. A broad peak was found in the rain study, see Fig. 3 \cite{Peters_Neelin2006} centred below the value of the water vapour at which the precipitation (order parameter) picks up. An even broader peak was found for the residence times in the brain study, see Fig. 3 panel E in \cite{Chialvo2012} and Fig. 10 in \cite{Scott2014}. In brain study the peak appears roughly to be centred in the vicinity where the order parameter starts to assume a significant non-zero value. 
\\Fig. \ref{Dist_Resi} contains the distribution of residence times and the distribution of the order parameter for $L=100$. 

  It is very demanding numerically to obtain good statistics for the joint distributions depicted in Fig. \ref{Order_para_Suscep}, hence the small systems size. To be able to do bigger systems we turn to one point measures below, but before that we want to point out that the distributions in Fig. \ref{Order_para_Suscep} do suggest that the onset of something like critical behaviour occurs in a broad range. 

Next we look at size scaling and are in particular interested in how the qualitative phenomenology of the FFM compares with the observations for rain and brain as we increase the system size and how the peak of the residence time distribution (left panel Fig. \ref{Dist_Resi}) changes as we increase the linear size of the system.  

We find that the peak of the residence time is to a very good approximation Gaussian with a skewness essentially equal to zero.   

\begin{figure}[!h]
\centering
\includegraphics[width=13cm,height=10cm]{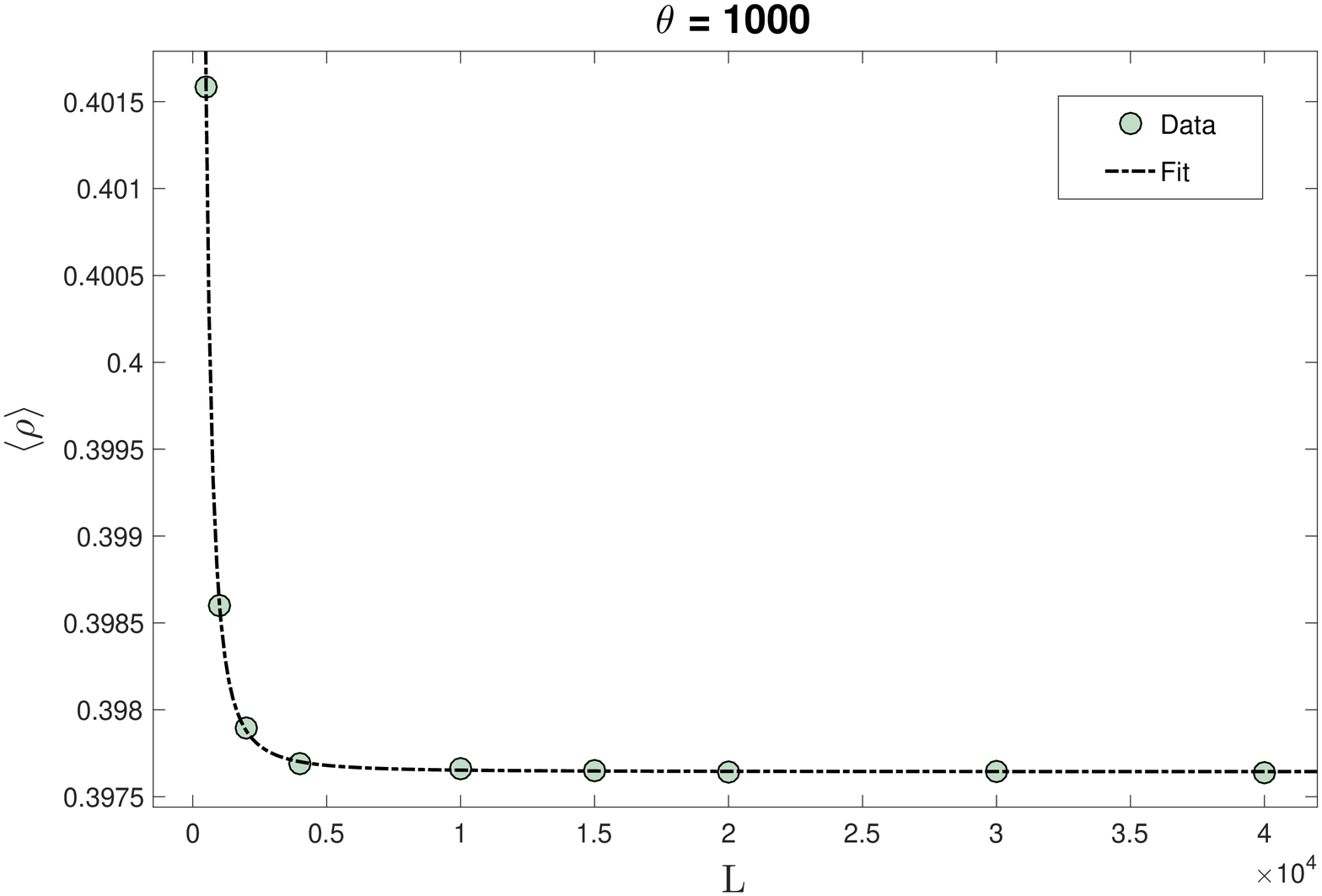}\\
  
\caption{Double logarithmic plot of the mean value $\langle \rho\rangle$  of the density of trees as function $L$}
\label{mean_rho}
\end{figure}

\begin{figure}[!h]
\centering
\includegraphics[width=13cm,height=10cm]{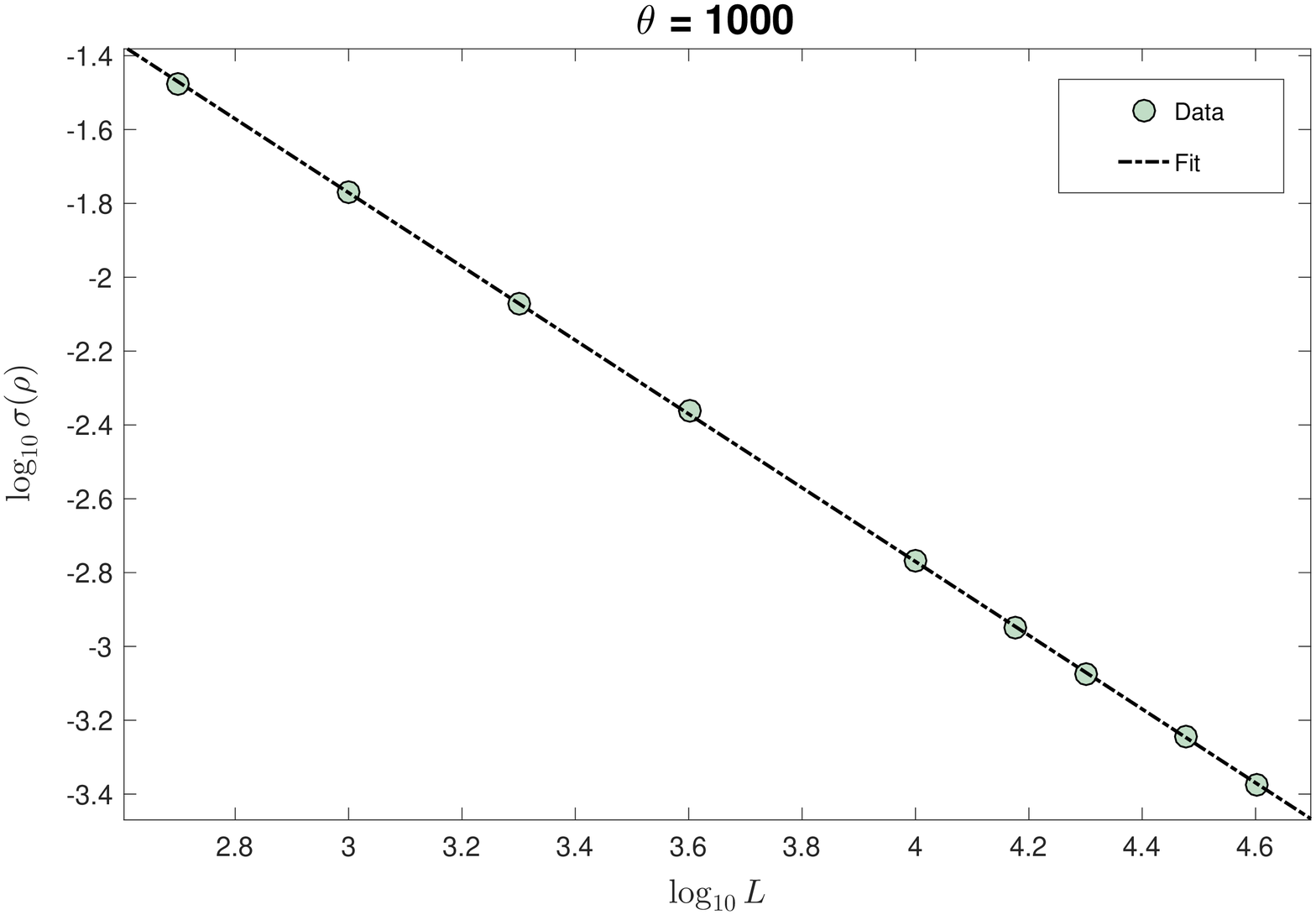}\\
  
\caption{Double logarithmic plot of the width $\sigma$ of the peak in the residence time distribution as function $L$}
\label{std(rho)}
\end{figure}

The average tree density as function of system size approaches rapidly the its asymptotic value, as see in Fig. \ref{mean_rho}. The behaviour is well fitted by $\rho(L)=\rho_\infty+bL^{-a} $ and $\rho_\infty=0.3976\pm 0.0001$.

The standard deviation of the residence time distribution is shown as function of system size in Fig. \ref{std(rho)} and we find that $\sigma(L)$ is well fitted by $\sigma(L)= 16.82/L$. This behaviour is consistent with the suggestion in \cite{Pruessner2002} that the variance may decreases as $L^2$, corresponding to the number of trees on the lattice effectively being uncorrelated so that $\rho=\sum_{i=1}^{L^2}x_i$ can be see as a sum of uncorrelated variable $x_i$, where $x_i=1$ i the site $i$ is occupied by a tree and $x_i=0$ otherwise.

\begin{table}[]
\centering
\caption{The total probability that the system is at a tree density more than one standard deviation from the average}
\label{my-label}
\begin{tabular}{|c|c|}
\hline
\rowcolor[HTML]{BBDAFF}
\textbf{L} & \textbf{Support} \\ \hline
\rowcolor[HTML]{ECF4FF}
20000      & 0.327            \\ \hline
\rowcolor[HTML]{ECF4FF}
15000      & 0.309            \\ \hline
\rowcolor[HTML]{ECF4FF}
10000      & 0.320            \\ \hline
\rowcolor[HTML]{ECF4FF}
5000       & 0.318            \\ \hline
\end{tabular}
\label{Table1}
\end{table}

\begin{figure}[!h]
\centering
\includegraphics[width=13cm,height=10cm]{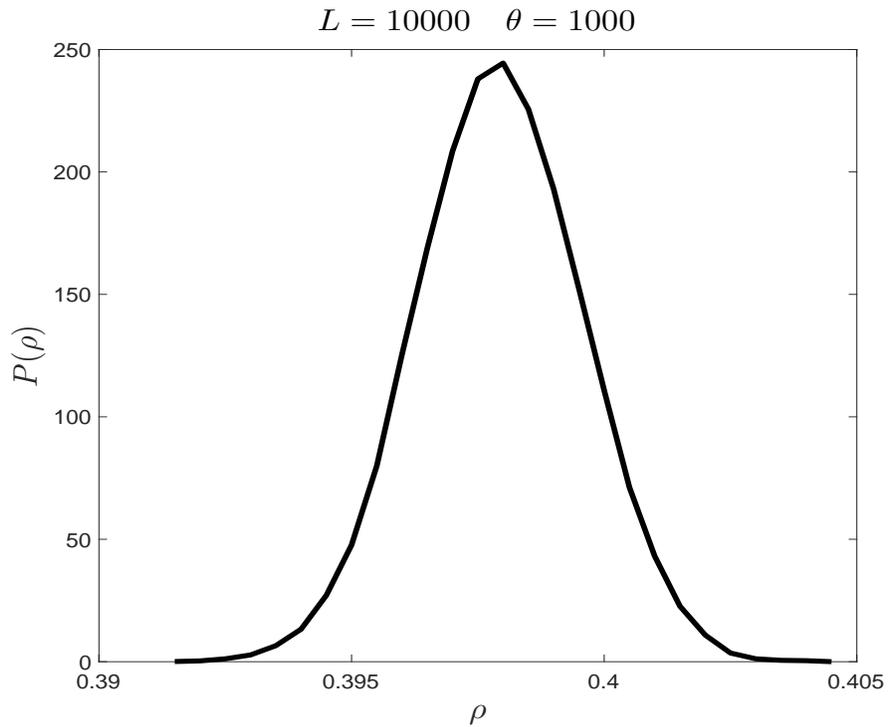}\\
  \caption{The residence time distribution of the tree densities for $L=10,000$ and $\theta=1000$. Same distribution as shown in the left panel of Fig. \ref{Dist_Resi} for $L=100$.}
\label{Large_resi}
\end{figure}

Table 1 lists the probability to find the system at a density one standard deviation away from the average for different system sizes. Gaussian behaviour with a probability for this deviation of about 32\% is found for all system sizes. The Gaussian form is also directly evident from the plot of the distribution of residence times in Fig. \ref{Large_resi}.

\begin{figure}[!h]
\centering
\includegraphics[width=13cm,height=10cm]{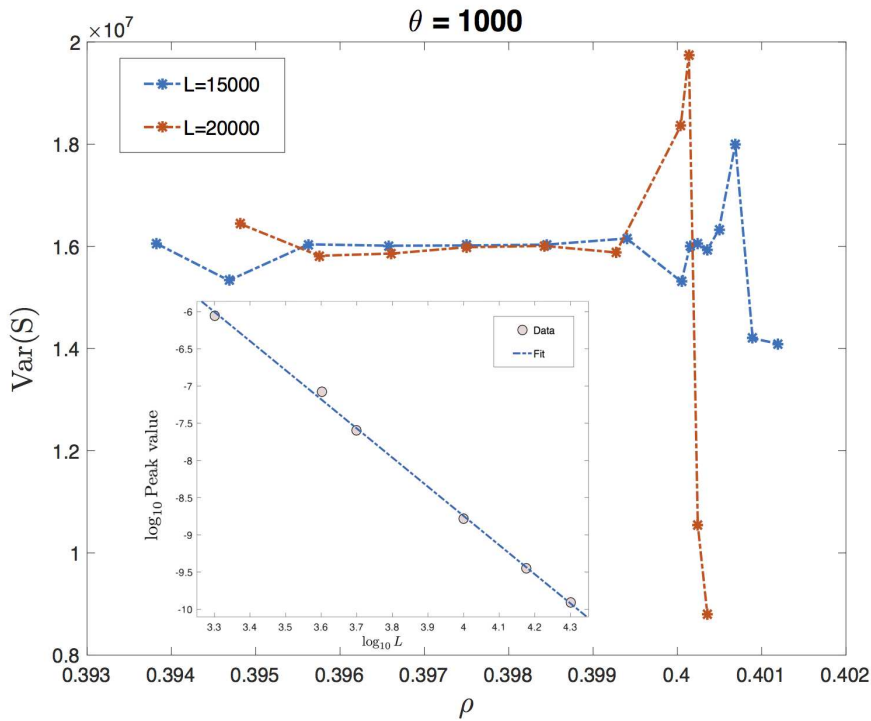}\\
  \caption{The variance of the distribution of cluster sizes. The insert is a log-log plot of the size dependence of the \textit{normalised} hight of the peak.  The slope of the line is $-3.917\pm 0.213$. The peak of the variance of the cluster size distribution (non-normalised) is obtained by multiply by $L^4$ and will accordingly diverge very slowly like $L^a$ with $a\simeq0.083$.}
\label{susceptible}
\end{figure}

\begin{figure}[!h]
\centering
\includegraphics[width=13cm,height=10cm]{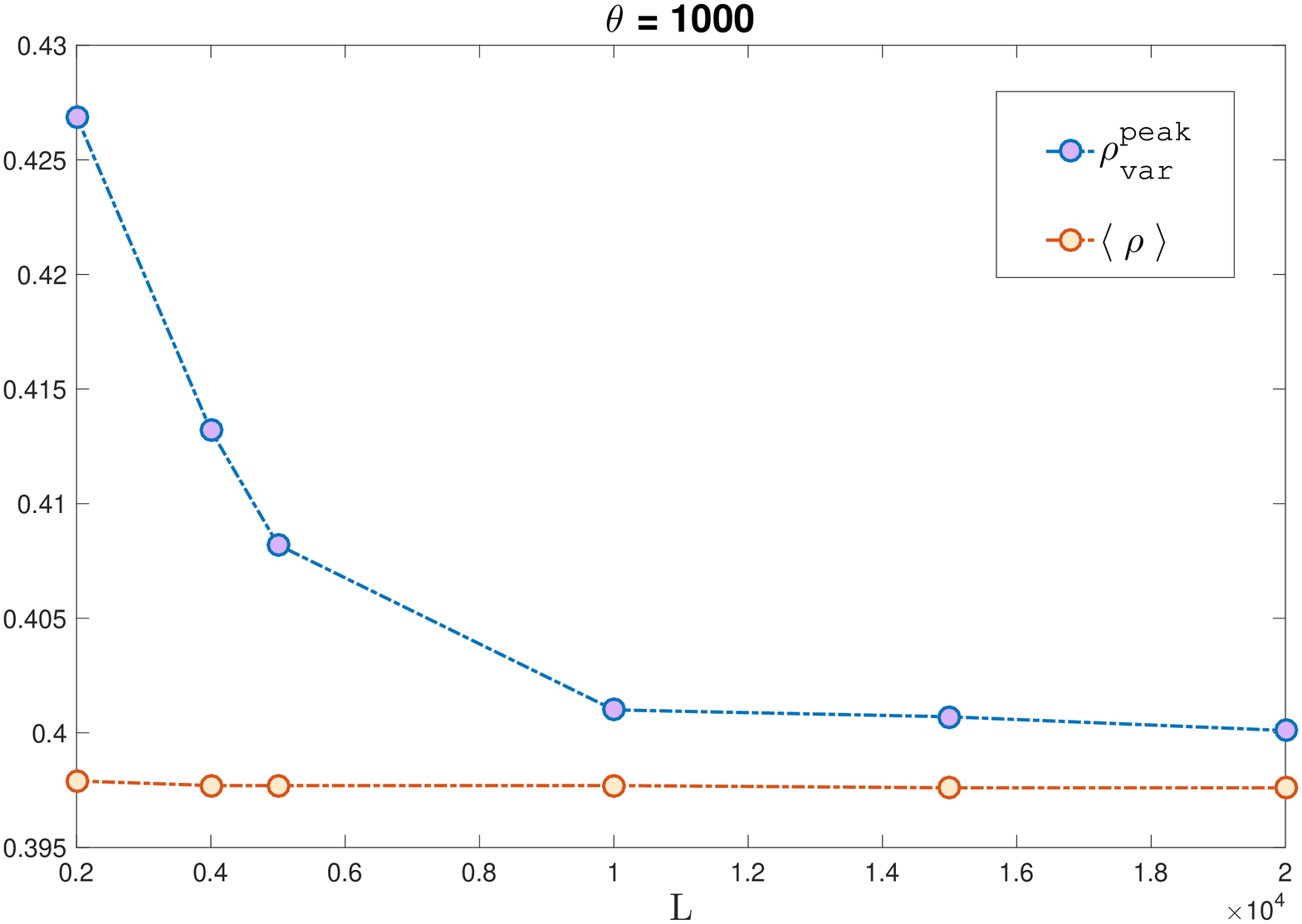}\\
  \caption{The location of the peak in the variance $\rho_{var}^{peak}$ and the mean tree density $\langle  \rho\rangle$ as function of linear system size $L$. Note $\langle  \rho\rangle$ is also the location of the peak in the residence time distribution.}
\label{approach}
\end{figure}

In Fig. \ref{susceptible} we show the variance of the cluster size distribution for different system sizes $L=15,000$ and $20,000$. The peak of the distribution is in both cases located at tree densities above the value at which the residence time peaks. The peak in the residence time is located at $\langle \rho\rangle$ (see Fig. \ref{mean_rho}). In Fig. \ref{approach} we show the location of the peak in the variance $\rho_{var}^{peak}(L)$ as function of system size together with $\langle \rho\rangle(L)$. For all system sizes the peak in the variance is above the peak in the residence time. Unfortunately it is difficult to conclude how  $\rho_{var}^{peak}(L)$ and  $\langle \rho\rangle(L)$ behaves as $L$ goes to infinity. 

The tree density at the location of the peak in the residence time, given by the average $\langle \rho\rangle$, is roughly located in the region where the order parameter picks up as indicated by Fig. \ref{Order_para_Suscep}. The pick up of the order parameter obviously occurs over a very broad range of values of tree densities and in principle this range will depend on the size of the system. To get high precision statistics for this is difficult but our simulations indicate that the order parameter picks up over an extended range of tree densities even for large system sizes. The broad pick up region and the same relative position of peaks is observed in the rain and brain studies, for rain see Fig. 1 and 3 in \cite{Peters_Neelin2006} and for brain see Fig. 3 panel E in \cite{Chialvo2012} and  Fig. 10 in \cite{Scott2014}.

\begin{figure}[!h]
\centering
\includegraphics[width=13cm,height=10cm]{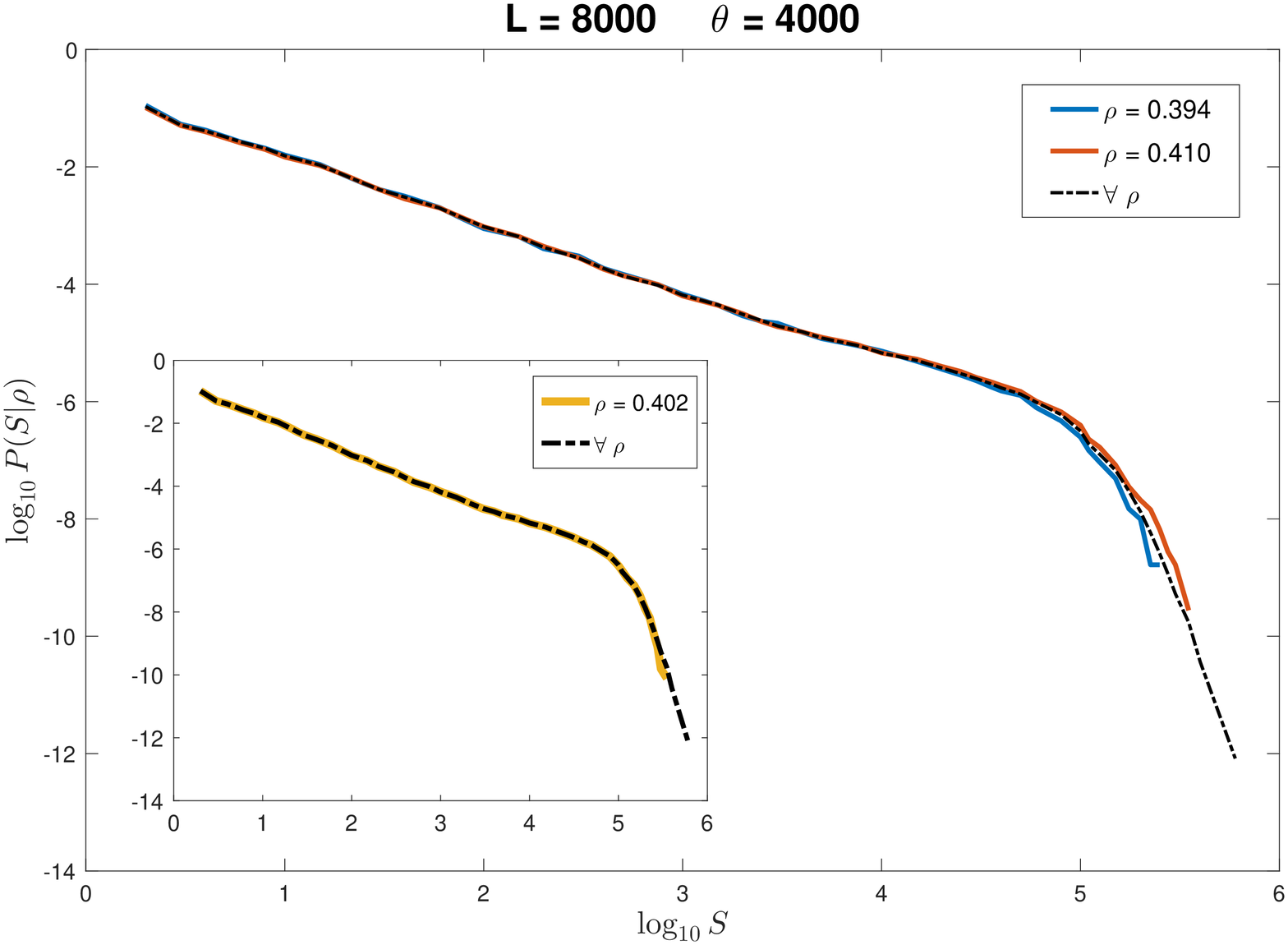}\\
  
\caption{Double logarithmic plot of the distribution of cluster sizes conditioned on given densities of trees}
\label{Cluster_Condi}
\end{figure}

In SOC studies observed tendencies to power law behaviour of event size distributions have been taken as a signature of "critical" behaviour \cite{Jensen1998,Gunnar_Book} ever since the original BTW paper \cite{BTW1987}. The rain and brain study we are inspired by likewise exhibit power law like distribution of event sizes. This was reported for rain in \cite{Peters2002} and for the brain activity see Fig. 3 panel F in \cite{Chialvo2012} and Fig. 9 in \cite{Scott2014}. In the FFM it is, of course, the distribution of burned tree clusters which exhibits this power law like behaviour and has been studied extensively in the literature, see e.g. \cite{Gunnar_Book} and for careful studies of deviation from proper power law behaviour see  \cite{Grassberger1993,Pruessner2002,Grassberger2002}. We want to add a comment to this debate seen in the light of the studies of rain and brain activity. In both studies it is at least tacitly assumed and in fact made explicit in \cite{Chialvo2012} that the control parameter fluctuates between low values at which the system can be seen as subcritical to high values where the system finds itself in a supercritical state. To understand how the FFM behaves in this respect we plot in Fig. \ref{Cluster_Condi} log-log plots of the distribution of sizes of tree clusters conditioned on specific values of the density of trees together with a plot of cluster sampled over all densities. The latter plot is the one considered previously in the literature and our graph for all tree densities exhibits the same kind of approximate power law behaviour as, e.g., seen in \cite{Pruessner2002}.

We note that the cluster size distribution sampled for the tree density $\rho=0.394$ is lightly below the overall density (see Fig. \ref{mean_rho}) and the density $\rho=0.410$ is slightly above the overall density. These two conditioned cluster size distributions could be said to exhibit under and over critical behaviour in the sense that their tail is a little bit above and a little bit below the overall cluster size distribution. A behaviour reminiscent of the one reported for brain activity in Fig. 3 panel F in \cite{Chialvo2012}. And, remarkably, the cluster size distribution sampled for $\rho=0.402$ coincides with the overall distribution (see insert in Fig. \ref{Cluster_Condi}). This behaviour might inspire one to think of the FFM driving itself around in a region of a critical point located at $\rho=0.402$ and while doing so mixing sub and super critical behaviour together. 

To what extend this scenario describes exactly the dynamics and statistics of the FFM is not entirely clear.  We do not know if, with increasing system size, scaling towards exact power law behaviour is happening for the cluster distribution conditioned for some  appropriately value $\rho(L)$. To answer this question we would need to do extensive size scaling of the conditioned cluster size distributions, which, unfortunately, is numerically very demanding.



\section{Discussion}
It is very difficult numerically to obtain with certainty the residence time distribution, the order parameter and the variance of the cluster size distribution. One will expect that if the residence time distribution narrows down to a delta peak positioned at the density at which the variance diverges and the order parameter picks up, then the infinite FFM would exhibit a critical state like the one we know in equilibrium systems at a $2^{nd}$ order phase transition. 

However, it may very well be that the dynamics for all finite systems sizes move the control parameter sufficiently around to complicate scaling. An extra complication consist in disentangling the scaling as the dissipation rate $f/p$ is taken to zero from the finite system size effects\cite{Pruessner2002}.

Though we do suggest that the FFM might be a useful laboratory to investigate behaviour seen in real "SOC" systems such as rain and brain, we also acknowledge that the origin of the broad cluster size distribution is still far from evident. The model self-organise to a density $\rho\simeq 0.4$ which is far below the critical density for percolation $\rho_c^{perc}\simeq0.6$. So it is not at all obvious that the tendency to power laws, or scale free behaviour, see in the FFM is induced by mechanisms familiar from percolation. But given the similarity between the FFM phenomenology and the rain and brain analysis, understanding how approximative power law-like behaviour arises in the FFM, if not related to percolation, of course makes a better understanding more urgent. 
  





\vspace{6pt} 


\noindent {\bf Acknowledgement}\\
LP gratefully acknowledges an EPSRC-Roth scholarship from the Department of Mathematics at Imperial College London. HJJ thanks Si Chen for interaction at the early stage of this project and is delighted to acknowledge a very helpful conversation with Dante Chialvo.\\

\noindent {\bf Author contributions}\\
{Both authors developed the conceptual ideas behind the research. L.P. performed the simulations. Both authors discussed the simulation results and jointly wrote the paper.}\\

\noindent {\bf Conflict of interests}\\
{The authors declare no conflict of interest.} 





\bibliographystyle{mdpi}



\end{document}